\documentclass[reprint,superscriptaddress,twocolumn, amsmath,amssymb,prb]{revtex4}

\usepackage{graphicx} % Include figure files
\usepackage{dcolumn}  % Align table columns on decimal point
\usepackage{bm}       % Bold math
\usepackage{xcolor}   % Allow for colored notes
\usepackage{epstopdf}
\begin{document}

\preprint{APS/123-QED}

\title{Anisotropic transport and quantum oscillations in the quasi-one-dimensional TaNiTe$_5$: Evidence for the nontrivial band topology}

\author{C. Q. Xu}
\affiliation{Department of Applied Physics, Zhejiang University of Technology, Hangzhou 310023, China}
\affiliation{Department of Physics and Astronomy, Michigan State University, East Lansing, Michigan 48824-2320, USA}
\author{Y. Liu}
\affiliation{Department of Applied Physics, Zhejiang University of Technology, Hangzhou 310023, China}
\author{P. G. Cai}
\affiliation{Department of Applied Physics, Zhejiang University of Technology, Hangzhou 310023, China}
\author{B. Li}
\affiliation{New Energy Technology Engineering Laboratory of Jiangsu Province and School of Science, Nanjing University of Posts and Telecommunications, Nanjing 210023, China}
\affiliation{National Laboratory of Solid State Microstructures, Nanjing University, Nanjing 210093, China}
\author{W. H. Jiao}
\affiliation{Department of Physics, Zhejiang University of Science and Technology, Hangzhou 310023, China}
\author{Y. L. Li}
\affiliation{Key Laboratory of Artificial Structures and Quantum Control (Ministry of Education), Shenyang National Laboratory for Materials
Science, School of Physics and Astronomy, Shanghai Jiao Tong University, Shanghai 200240, China}
\author{J. Y. Zhang}
\affiliation{Department of Physics, Changshu Institute of Technology, Changshu 215500, China}
\author{W. Zhou}
\affiliation{Department of Physics, Changshu Institute of Technology, Changshu 215500, China}
\author{B. Qian}
\affiliation{Department of Physics, Changshu Institute of Technology, Changshu 215500, China}
\author{X. F. Jiang}
\affiliation{Department of Physics, Changshu Institute of Technology, Changshu 215500, China}
\author{Z. X. Shi}
\affiliation{School of Physics and Key Laboratory of MEMS of the Ministry of Education, Southeast University, Nanjing 211189, China}

\author{R. Sankar}
\affiliation{Institute of Physics, Academia Sinica, Nankang, Taipei, 11529, Taiwan}
\author{J. L. Zhang}
\affiliation{High Magnetic Field Laboratory, Chinese Academy of Sciences, Hefei 230031, China}
\author{F. Yang}
\affiliation{Wuhan National High Magnetic Field Center, School of Physics, Huazhong University of Science and Technology, Wuhan, 430074, China}
\author{Zengwei Zhu}
\affiliation{Wuhan National High Magnetic Field Center, School of Physics, Huazhong University of Science and Technology, Wuhan, 430074, China}
\author{P. K. Biswas}
\affiliation{ISIS Pulsed Neutron and Muon Source, STFC Rutherford Appleton Laboratory, Harwell Campus, Didcot, Oxfordshire OX11 0QX, United Kingdom}
\author{Dong Qian}
\affiliation{Key Laboratory of Artificial Structures and Quantum Control (Ministry of Education), Shenyang National Laboratory for Materials
Science, School of Physics and Astronomy, Shanghai Jiao Tong University, Shanghai 200240, China}
\affiliation{Tsung-Dao Lee Institute, Shanghai Jiao Tong University, Shanghai 200240, China}

\author{X. Ke}
\email{kexiangl@msu.edu}
\affiliation{Department of Physics and Astronomy, Michigan State University, East Lansing, Michigan 48824-2320, USA}

\author{Xiaofeng Xu}
\email{xuxiaofeng@zjut.edu.cn}
\affiliation{Department of Applied Physics, Zhejiang University of Technology, Hangzhou 310023, China}

\date{\today}

\begin{abstract}
The past decade has witnessed the burgeoning discovery of a variety of topological states of matter with distinct nontrivial band topologies. Thus far, most of materials studied possess two-dimensional or three-dimensional electronic structures, with only a few exceptions that host quasi-one-dimensional (quasi-1D) topological electronic properties. Here we present the clear-cut evidence for Dirac fermions in the quasi-1D telluride TaNiTe$_5$. We show that its transport behaviors are highly anisotropic and we observe nontrivial Berry phases via the quantum oscillation measurements. The nontrivial band topology is further corroborated by first-principles calculations. Our results may help to guide the future quest for topological states in this new family of quasi-1D ternary chalcogenides.
\end{abstract}

\maketitle

\section{Introduction}
The interplay between dimensionality and electronic quantum states has been a fundamental theme of condensed matter physics for many decades. As a prominent example, it is well established that when electrons are spatially confined to a one-dimensional (1D) chain, interactions will drive the system into
the 1D Luttinger-liquid regime with a characteristic feature of spin-charge separation\cite{S. Tomonaga,J. M. Luttinger}. Recently, the exploration of solids with various nontrivial band topologies has become one of the heatedly pursued topics. Surprisingly, the vast majority of topological materials identified to date are electronically two-dimensional or three-dimensional in essence, with much fewer one-dimensional counterparts known to exist.

The quasi-1D topological material candidates reported thus far mainly involve the halogen compounds $\beta$-Bi$_4$X$_4$(X = Br, I)\cite{Bi4I4 nature material,Bi4I4 Bi4Br4,alfa-beta Bi4I4}, $\alpha$-Bi$_4$I$_4$\cite{alfa-beta Bi4I4,alfa-Bi4I4}, (TaSe$_{4}$)$_{2}$I\cite{J. Gooth TaSeI,TaSeI-arxiv 2,TaSeI-arxiv 3}, the Weyl semimetal (Ta, Nb)IrTe$_4$\cite{TaIrTe4-1,TaIrTe4-2,NbIrTe4 Zhou,NbIrTe4 PRB}, and the theoretically proposed molybdenum chalcogenides A$_2$Mo$_6$X$_6$ (A = Alkani, In, Tl; X = chalcogen)\cite{Tl2Mo6Se6,S. Mitra Tl2Mo6Se6,Shin-Ming Huang Tl2Mo6Se6}, etc. Among them, $\beta$-Bi$_4$I$_4$ has been experimentally verified as a weak topological insulator, i.e., the 3D stacking of 2D quantum spin Hall (QSH) states, where topological surface states emerge only on the side surfaces but not on the top and bottom surfaces\cite{alfa-beta Bi4I4}. (TaSe$_{4}$)$_{2}$I was theoretically proposed and experimentally confirmed to be a type-III Weyl semimetal with larger chiral charges which can support four-fold helicoidal surface states with significantly long Fermi arcs\cite{TaSeI-arxiv 2}.  TaIrTe$_4$ is a type-II Weyl semimetal with the minimum number of four Weyl points, which however are located above the Fermi level, making it challenging for real applications. The molybdenum chalcogenides A$_2$Mo$_6$X$_6$ were recently proposed to be a topological superconductor hosting cubic Dirac Fermions, which shows linear band crossing along one principal axis but cubic dispersion in the plane perpendicular to it. Nevertheless, due to the needle-shaped morphography of the as-grown samples, experimental validation of this type of topological states proves to be extremely challenging.

In this article, we report nontrivial topological electronic properties in the chain-containing ternary telluride TaNiTe$_5$ via comprehensive magnetization and transport measurements. The one-dimensional NiTe$_2$ chains are parallel to the crystallographic $a$-axis. As a result, this material exhibits highly anisotropic transport behaviors with $\rho_{a}$: $\rho_{b}$: $\rho_{c}$ $\sim$ 1 : 16 : 7 at 320 K. The quantum oscillations in both isothermal magnetization and magnetotransport measurements reveal light effective quasiparticle masses and nontrivial Berry phases. The nontrivial band topology is further supported by \textit{ab initio} calculations. Our results establish that TaNiTe$_5$ represents a rare example of quasi-1D topological materials and is suitable for the study of interplay between dimensionality and band topology.

\section{Experimental details}

TaNiTe$_{5}$ single crystals were synthesized by the self-flux method. High-purity tantalum pieces, nickel shots and tellurium ingots were mixed with a molar ratio of 1 : 1 : 10 and sealed in an evacuated quartz tube. The quartz tube was then loaded into a box furnace and heated up to 1273 K quickly and kept at this temperature for 24 hours to ensure complete melting. The furnace was then slowly cooled down to 773 K with a rate of 5 K/h. After centrifugation at 773 K and fast-cooling to room temperature, shiny TaNiTe$_{5}$ single crystals in millimetre-long needles with an aspect ratio of $\sim$ 1:10 (as shown in Fig. 1(c)) were obtained.

We checked the crystal quality and determined the structure of the as-grown samples using a four-circle X-ray diffractometer. The electrical resistivity and magneto-transport measurements were conducted using the Physical Property Measurement System (PPMS-9), and the magnetic susceptibility was measured using the Magnetic Property Measurement System (MPMS), both from Quantum Design. Magnetotransport measurements in the pulsed magnetic field up to 50 T were measured in Wuhan National High Magnetic Field Center, and Torque measurements were performed in the High Magnetic Field Laboratory in Hefei.

We also performed density functional theory (DFT) calculations by means of the full-potential linearized augmented plane wave method implemented in the WIEN2K package~\cite{Wien2k} to obtain the electronic structure of TaNiTe$_{5}$. The generalized gradient approximation (GGA) presented by Wu and Cohen~\cite{GGA} was used for the exchange-correlation energy. This GGA is a nonempirical approximation that gives a significant improvement of calculations for lattice constants and crystal structures. Relativistic effects and spin-orbit coupling are included in all calculations. To further identify the topological properties, a tight-binding model based on maximally localized Wannier functions~\cite{Mostofi2014An,Wu2017WannierTools} was constructed to reproduce the bulk band structure including spin-orbit coupling of Ta $d$, Ni $d$ and Te $p$ orbits. Fermi surface was calculated on a dense $35\times35\times9$ $k$-point mesh in the full Brillouin zone. The extremal-orbit quantum oscillation frequencies of the calculated Fermi surfaces were analyzed using SKEAF (Supercell K-space Extremal Area Finder)~\cite{skeaf}.

\section{Results}

TaNiTe$_{5}$ crystallizes in the orthorhombic space group ($Cmcm$, No. 63) with the lattice parameters of $a$ = 3.6674(7){\AA}, $b$ = 13.172(3){\AA}, $c$ = 15.142(3){\AA}, and $\alpha = \beta = \gamma$ = $90^{\circ}$\cite{TaNiTe5}. As shown in Fig. 1, the structure consists of Ta bicapped trigonal prismatic chains which face-share with the adjacent Ni octahedral chains along the $a$-axis. Along the chain direction, two neighboring Ta atoms are connected via three Te atoms, and two neighboring Ni atoms are connected via two Te atoms\cite{MMTe5}. In contrast, the compound forms layer structure with large spacing between layers along the $b$-axis, as illustrated in Fig. 1(b). Fig. 1(c) shows the optic image of the crystals which exhibit a needle-like shape with cleavable flat surface perpendicular to the $b$-axis. Fig. 1(d) shows the temperature dependence of resistivity measured at zero field along three crystallographic axes. Evidently, this material exhibits quasi-1D transport behaviors with $\rho_{a}$: $\rho_{b}$: $\rho_{c}$ $\sim$ 1 : 16 : 7 at $T$ = 320 K and $\sim$ 1 : 10 : 12 at $T$ = 2 K.  Additionally, the resistivity closely follows linear-$T$ dependence from 320 K to 50 K and crosses over to a Fermi liquid behavior with a quadratic $T$ dependence below 50 K, as demonstrated in the inset of Fig. 1(d) for $\rho_{b}$ (data of $\rho_{a}$ and $\rho_{c}$ show similar dependence and are not shown here), suggesting that the electron-electron scattering dominates in the low-$T$ region.

\begin{figure}
\includegraphics[width=8cm]{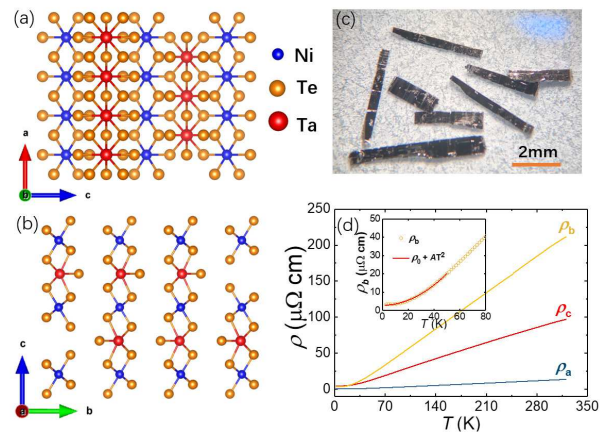}% Here is how to import EPS art
\caption{\label{fig1} Crystal structure of TaNiTe$_{5}$ viewed along the $b$-axis (a), and the $a$-axis (b). (c) The optical photo of some pieces of needle-like crystals. The longest direction is along the $a$-axis. (d) The temperature dependence of resistivity $\rho_{a}$, $\rho_{b}$, $\rho_{c}$ with the current applied along the $a$, $b$, $c$ axes, respectively. The inset shows the fit to the Fermi-liquid paradigm for a representative $\rho_{b}$.}
\end{figure}

Figure 2 displays the transverse magnetoresistance (MR) under different current and field-direction configurations. Interestingly, the MR of intrachain $\rho_{a}$ shows notable difference compared to that of interchain $\rho_{b}$ and $\rho_{c}$. Firstly, while $\rho_{a}$ exhibits metallic behavior down to the lowest temperature even under a field of 9 T, a field-induced metal-to-insulator like transition was observed near $T$ = 30 K for both $\rho_{b}$ and $\rho_{c}$ once $\mu_{0}H\geq$ 7 T, as presented in Fig. 2 (a, d, g). Secondly, in panels (b), (e) and (h), we plot the MR curves measured at fixed temperatures. As seen, MR is remarkably small for $\rho_{a}$ and it tends to saturate with a low magnitude of $\sim$ 26\% at 2 K and 9 T. By contrast, $\rho_{b}$ and $\rho_{c}$ reaches a MR as high as 700$\%$ and 500$\%$ respectively without saturation at the same settings. These anisotropic transports signify the large anisotropic Fermi surfaces and the associated electron life time, reflecting its quasi-1D electronic structure. Finally, Fig. 2(c, f, i) show the Kohler's plots for these three configurations. It is evident that, for $\rho_{b}$ and $\rho_{c}$, the Kohler's rule is well obeyed. However, for $\rho_{a}$, it is moderately violated, presumably due to the anisotropy in its carrier scattering time \cite{PdSn4,N.Luo}. Overall, these anisotropic magneto-transport behaviors also confirm the anisotropic electronic properties in this quasi-1D TaNiTe$_{5}$.

\begin{figure}
\includegraphics[width=9.5cm]{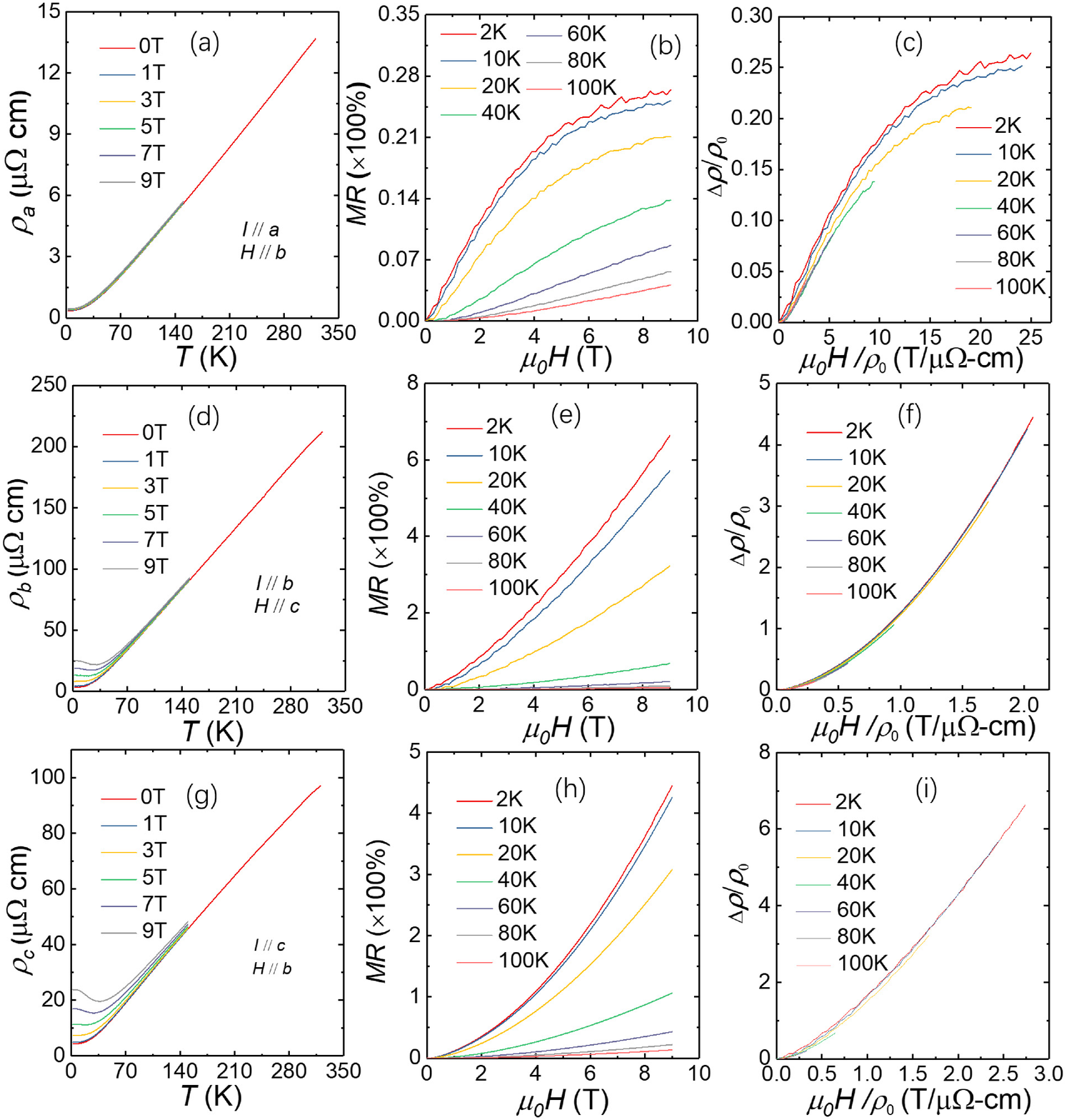}% Here is how to import EPS art
\caption{\label{fig2} (a-c) The temperature sweeps of $\rho_{a}$ under various fields, the MR at fixed temperatures, and the Kohler's plot under the configuration of $I \parallel a$ and $H \parallel b$, respectively. (d-f) The same for $I \parallel b$ and $H \parallel c$. (g-i) The same for $I \parallel c$ and $H \parallel b$.}
\end{figure}

\begin{table*}[t]
    \centering
    \caption{Parameters obtained from the fitting of dHvA and SdH oscillations}
    \begin{tabular}{cccccccccccccccc}
        \hline\hline
         & & &$H$ $\parallel a$  &  & & &  & &  $H$ $\parallel b$ \\

       & (unit) & & dHvA   & & & & dHvA &  & & & & SdH \\
       \hline
         &  & $\alpha$ &  $\beta$ & & & $F_{1}$ & $F_{2}$  & $F_{3}$ & $F_{4}$ & & $F_{1}^{'}$ & $F_{2}^{'}$ &$F_{*}^{'}$ & $\delta$ & $F_{4}^{'}$ \\
        $F$ &($T$)  & 63 & 251 &  & & 56 & 163 &   231 & 763 & & 57 & 155 & 312& 471& 731 \\
        $A$ &($nm^{-2}$)  & 0.600 & 2.391 &  & & 0.533 & 1.552 &   2.200 & 7.267 & & 0.543 & 1.476 &2.972& 4.486& 6.962 \\
        $m/m_{e}$ &  & 0.190 & 0.217 &  & & 0.169 & 0.173 &   0.180 & 0.295 & &  & 0.225 & 0.311& 0.188& 0.291 \\
        $T_{D}$ &($K$)& 4.93 & 6.24 &  & & 5.26 & 5.97 &   7.17 & 6.34 & &  &  & & &  \\
        $\phi_{B}$ & ($\delta=-\frac{1}{8}$) & 1.497$\pi$ & 1.972$\pi$ &  & & 1.502$\pi$ & 0.870$\pi$ &   1.212$\pi$ & 0.730$\pi$ & &  &  & & &  \\
         & ($0$) & 1.247$\pi$ & 1.722$\pi$ & &  & 1.252$\pi$ & 0.620$\pi$ &   0.962$\pi$ & 0.480$\pi$ & &  &  & & &  \\
         & ($\delta = \frac{1}{8}$) & 0.997$\pi$ & 1.472$\pi$ &  & & 1.002$\pi$ & 0.370$\pi$ &   0.712$\pi$ & 0.230$\pi$ & &  &  & & &  \\
        \hline\hline
    \end{tabular}

\end{table*}

In order to gain more insights on the electronic structure of TaNiTe$_{5}$, we measured the quantum oscillations of its isothermal magnetization on single crystals with the magnetic field applied along two different crystallographic axes, i.e., $H\parallel a$ and $H\parallel b$. The insets of Fig. 3(a) and 3(b) present the isothermal magnetization $M(H)$ curves at $T$ = 2 K with the magnetic field along the $a$ and $b$ axes, respectively, where de Haas-van Alphen oscillations (dHvA) are clearly observable. Remarkably, the susceptibility shows anisotropic, diamagnetic signals over the whole field range with its anisotropic ratio ($\chi_b/\chi_a$) of $\sim$ 5 at $\mu_{0}$H = 7 T. Such anisotropy is commonly observed in layered materials and can be attributed to the anisotropic Lande-$g$ factors\cite{Terasaki92,Pr124 spin flop}. The oscillatory component becomes more visible after subtracting the diamagnetic background, as shown in Fig. 3(a) and 3(b). In Fig. 3(c) and 3(d), we plot the fast Fourier transform (FFT) analysis of $\Delta M$. Interestingly, only two fundamental frequencies at $F_{\alpha}= 63 $ T, $F_{\beta}= 251 $ T are obtained for the magnetic field applied along the $a$ axis; in contrast, four fundamental frequencies at $F_{1}= 56 $ T, $F_{2}= 163 $ T, $F_{3}= 231 $ T, $F_{4}= 763 $ T are clearly observed with the field along the $b$-axis. This again indicates that the Fermi surface morphology is indeed very anisotropic. Based on the Onsager relation $F=(\hbar/2\pi e)A_{e}$, we extracted the extremal Fermi-surface cross-sectional area $A_{e}$ associated with each fundamental frequency within the Brillouin zone, as listed in Table I.

To proceed, we further performed the quantitative analysis of oscillatory data by the standard Lifshitz-Kosevich (LK) formula\cite{LK 1,LK 2,LK 3}.
 \begin{equation}
\begin{aligned}
\Delta M \varpropto - R_{D}R_{T}\sin\{2\pi[\frac{F}{B}+(\frac{1}{2}-\phi)]\}
\end{aligned}
\end{equation}

\noindent where $R_{D}$ is the Dingle damping term, $R_{T}$ is the thermal damping factor. $R_{D} = \exp(2\pi^{2}k_{B}T_{D}m^{\ast}/eB\hbar)$, and $R_{T} = \frac{2\pi^{2}k_{B}Tm^{\ast}/eB\hbar}{sinh(2\pi^{2}k_{B}Tm^{\ast}/eB\hbar)}$, $\phi = \phi_{B}/2\pi - \delta$, where $k_{B}$ is the Boltzmann constant, $m^{\ast}$ is the effective electron mass, $T_{D}$ is the Dingle temperature, $\phi_{B}$ is the Berry phase and $\delta$ is an additional phase shift depending on the dimensionality of the Fermi surfaces, i.e., $\delta = 0$ for 2D Fermi surfaces, and $\delta = \pm\frac{1}{8}$ for 3D Fermi surfaces ($-\frac{1}{8}$ for the electron-like, $+ \frac{1}{8}$ for the hole-like, respectively). Therefore, one can determine the effective mass for each frequency by fitting the temperature dependence of the oscillatory components extracted from the FFT amplitude to $R_{T}$, as shown in Fig. 3(e) and 3(f). For $H \parallel a$, the effective electron masses corresponding to the two fundamental frequencies are $m_{\alpha} = 0.190 m_{0}$, $m_{\beta} = 0.217 m_{0}$ ($m_{0}$ is the free electron mass); for $H \parallel b$, the corresponding fits yield the effective masses of $m(F_{1}) = 0.169 m_{0}$, $m(F_{2}) = 0.173 m_{0}$, $m(F_{3}) = 0.180 m_{0}$, $m(F_{4}) = 0.295 m_{0}$. These small values of the effective mass for each frequency implies the presence of relativistic charge carriers. Besides, using the above fitted $m$, we can further fit $\Delta M$($1/B$) at a given temperature (e.g., $T$ = 2 K) to the LK formula directly to obtain the Berry phase for each band. As shown in Fig. 3(g) and (h), the oscillations can be perfectly fitted using the above formula and the fitting procedure gives the Berry phase and the Dingle temperature for each band, as listed in Table I for completeness. As noted, some of Berry phases are close to $\pi$, suggesting the non-trivial fermions in TaNiTe$_{5}$.

\begin{figure}
\includegraphics[width=8cm]{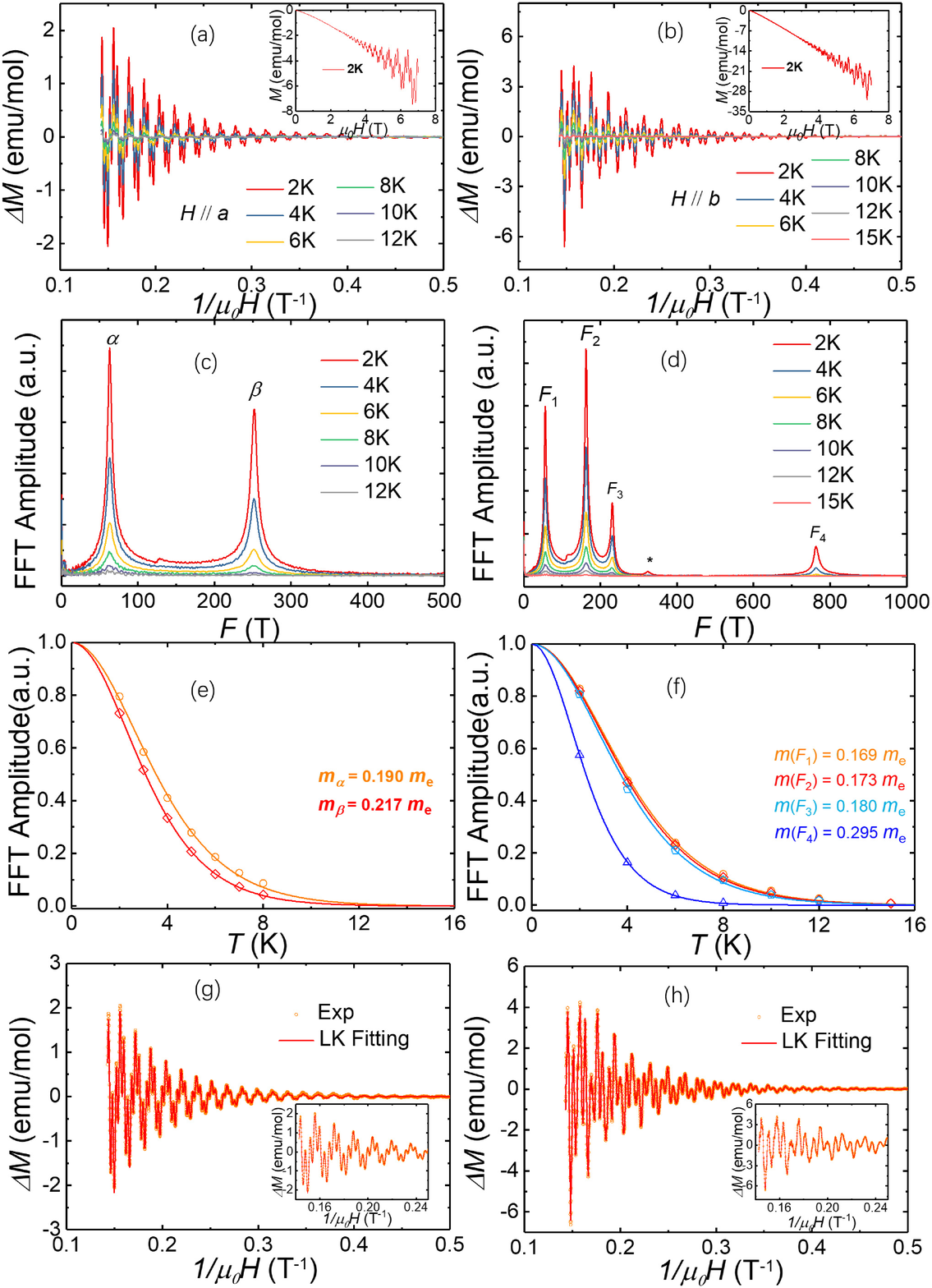}% Here is how to import EPS art
\caption{Oscillatory components $\Delta M$ vs $\mu_{0}H$ with $H \parallel a$ (a), $H \parallel b$ (b) at various temperatures. The insets of panel (a) and (b) are the raw data of $M-H$ at 2 K. (c) and (d) The FFT spectra of $\Delta M$ oscillations. (e) and (f) The temperature dependence of FFT amplitude. The solid lines are the LK fits for the effective mass. (g) and (h) The oscillatory components and the corresponding LK fits. The insets are the enlarged view to demonstrate the good fitting. }
\end{figure}

In addition, we probed the Shubnikov-de Haas (SdH) oscillations of the magnetoresistance by extending the magnetic field up to 50 T. Here, we measured the inter-chain resistivity $\rho_{c}$ instead of intra-chain $\rho_{a}$ because $\rho_{c}$ is the largest at low-$T$ such that it gives the best signal-to-noise level. As shown in Fig. 4(a), MR reaches a magnitude of 4300\% at $T$ = 2 K, $H$ = 50 T and the pronounced oscillations can been detected at high fields. After subtracting a smooth background, we extracted the oscillatory components, as illustrated in Fig. 4(b). From the FFT (Fig. 4(c)), we have derived five independent frequencies. Three of them ($F_{1}^{'}$, $F_{2}^{'}$, $F_{4}^{'}$) are comparable with those from dHvA and the other two ($F_{*}^{'}$ and $\delta$) seem to be the new frequencies. By employing the LK formula, the effective mass can be determined. Note that the $m$ observed from SdH is larger than that from dHvA, especially for $m(F_{2})$. This discrepancy between dHvA and SdH oscillations is often seen in low-dimensional materials such as the layered organic compounds and arises from the different mechanisms of dHvA and SdH oscillations\cite{HuJin}. It is known that SdH oscillations originate from the oscillating scattering rate and can thus be complicated by the detailed scattering procedures, imposed by lattice, impurity, inter/inner-Landau level scattering, etc.\cite{E. N. Adams,Gao wenshuai}. In contrast, dHvA effect is caused directly by the oscillations of electrons' free energy.

\begin{figure}
\includegraphics[width=9cm]{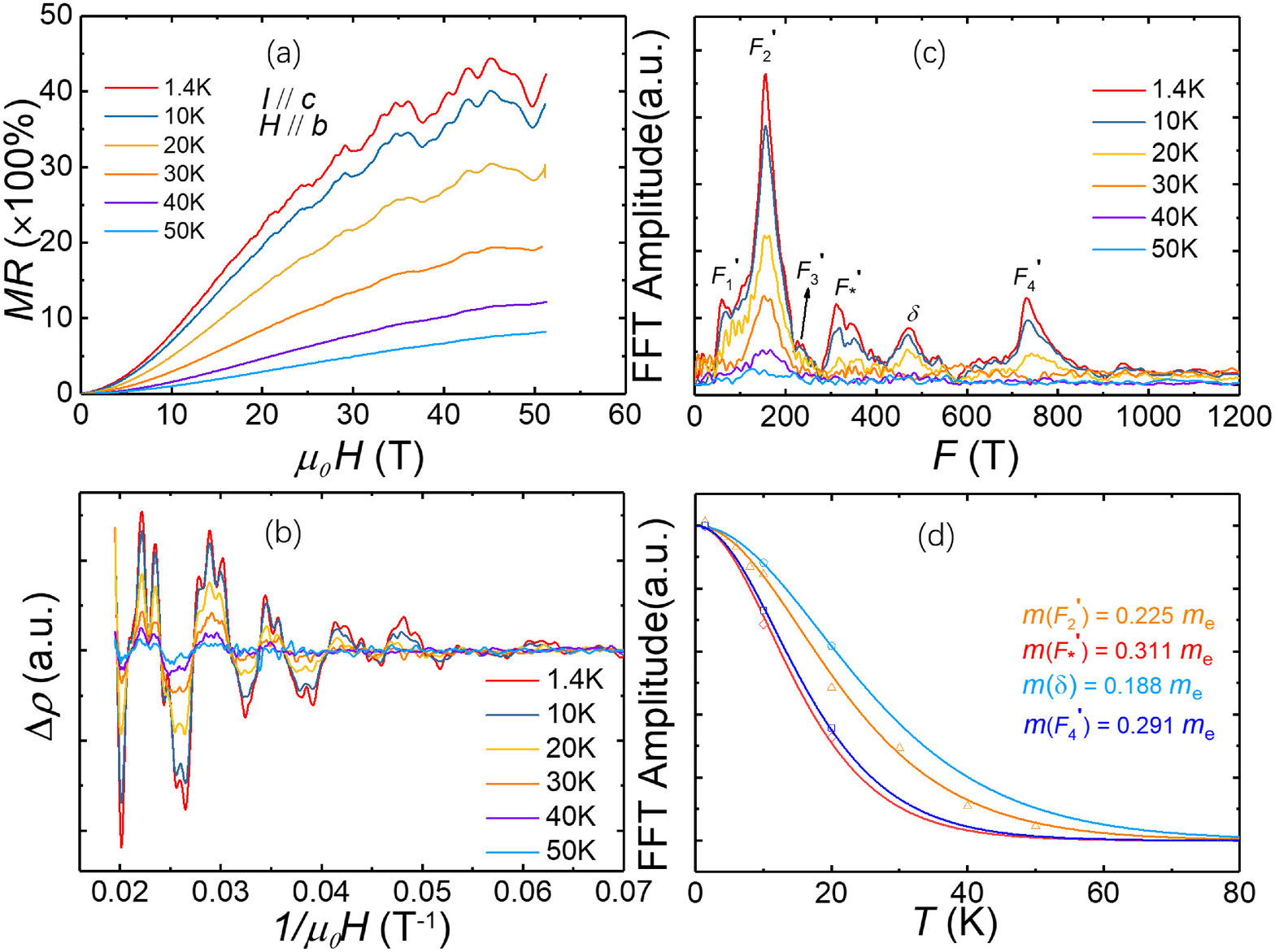}% Here is how to import EPS art
\caption{(a) MR at different temperatures with $I \parallel c$, $H \parallel b$  up to 50 Tesla. (b) The oscillatory components at different temperatures. (c) FFT spectra of the SdH oscillations at different temperatures. (d) The temperature dependence of FFT amplitude and its fits. Note that we can not obtain $m(F_{1}^{'})$ due to its noisy signals.}
\end{figure}

\begin{figure*}
\includegraphics[width=13cm]{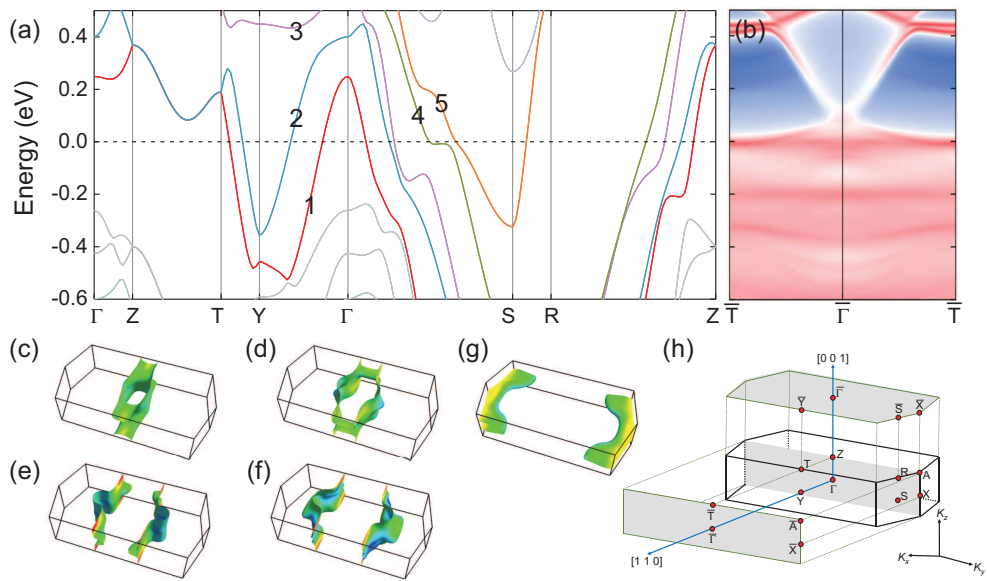}
\caption{\label{band} (Color online) (a) Calculated band structure for bulk TaNiTe$_5$ with SOC included. The bands crossing the Fermi level are marked by different colors and labeled by numbers. (b) Surface state spectrum of TaNiTe$_5$ in the presence of SOC along high symmetry $k$-line $\bar{T}-\bar{\Gamma}-\bar{T}$. The bright red lines denote the surface states. (c-g) 3D Fermi surfaces from band 1 to band 5 as labeled in (a) with Fermi velocities shadowed. (h) The bulk and surface projected Brillouin zone (BZ) for primitive cell with high-symmetry points specified.}
\end{figure*}

To better understand the electronic properties of TaNiTe$_5$, we have performed DFT calculations to obtain its electronic structure. The calculated electronic band structure is shown in Fig. \ref{band}(a), with five bands crossing the Fermi level. In common with the quasi-1D Tl$_2$Mo$_6$Se$_6$ compound \cite{Tl2Mo6Se6}, TaNiTe$_5$ also possesses the cubic Dirac crossings in the $\Gamma-Z-T-Y$ direction, which have the quasi-linear band crossing along $\Gamma-Z$ and $T-Y$ but the cubic dispersions along $Z-T$. The cubic Dirac crossings are located at high symmetry points $T$ (-0.5, 0.5, 0.5) and $Z$ (0, 0, 0.5), as shown in Fig. \ref{band}(a). The Fermi surface analysis has been performed to determine the band type in the full Brillouin zone. It is found that Te-derived 5$p$ orbital dominates around the Fermi level. The bulk Fermi surfaces with color-shadowed Fermi velocities are visualized in Fig. \ref{band}(c-g), which are comprised of five segments: band 1, band 2 are hole-dominated Fermi surfaces, and band 3, band 4 and band 5 are electron-dominated Fermi surfaces. The flatness of the Fermi surfaces indicates the quasi-1D feature expanding along the $b$ direction in real space, which responds to [110] direction in primitive reciprocal space. The calculated surface state spectrum along $\bar{T}-\bar{\Gamma}-\bar{T}$ is shown in Fig. \ref{band}(b). One can see significant bright red surface states along the slab direction forming Dirac-cone-like states at the slab center, which signify the topologically nontrivial characteristics of TaNiTe$_5$. The other piece of evidence for the topological non-triviality of TaNiTe$_5$ derives from its $\mathbb{Z}_2$ index. We calculated the $\mathbb{Z}_2$ topological number to further identify the topological properties. By using the Wilson loop technology~\cite{Wu2017WannierTools}, we obtained the $\mathbb{Z}_2$ topological number for TaNiTe$_5$, which is 1 for the $k_z$ = 0 plane, and zeros for other planes. The resultant topological index is (1, 000), whereby it makes a compelling case for the strong topological states in TaNiTe$_5$.

\section{Conclusion}

In summary, we reported the synthesis of a new quasi-1D ternary telluride TaNiTe$_5$ and identified its possible topologically nontrivial states by means of electrical transport, quantum oscillations and the first-principles calculations. Specifically, its one-dimensionality is manifested in its large anisotropy in the resistivity and magnetoresistance and the susceptibility. The topological carriers are supported by nontrivial Berry phase extracted from the quantum oscillations, the Dirac band dispersions and the nontrivial $\mathbb{Z}_2$ index from the calculations. The findings presented in this work shall certainly invoke more theoretical study and the experimental investigations, such as (spin-polarized) ARPES and scanning tunnel microscopy, to probe the nontrivial bulk and surface states proposed in this study.

\begin{acknowledgments}

The authors would like to thank C. M. J. Andrew, Ali Bangura, Xin Lu for useful discussions. This work is sponsored by the National Natural Science Foundation of China (Grant No. 11974061, No. U1732162, No. 11521404, No. U1632272, No. U1832147, No. U1932217). W. H. J thanks the financial support from Zhejiang
Provincial Natural Science Foundation of China (No. LY19A040002) and B. L. thanks NUPTSF (Grant No. NY219087, NY220038). R. S. acknowledges the financial support from the Ministry of Science and Technology in Taiwan under project number MOST-108-2112-M-001-049-MY2 and from Academia Sinica for the budget of AS-iMATE-109-13. X. Ke acknowledges the financial support from the start-ups at Michigan State University.

C. Q. Xu and Y. Liu contributed equally to this work.

\end{acknowledgments}

\appendix

\end{document}